\documentclass[12pt]{article}
\usepackage{amsxtra,amssymb,amsthm,amsmath,latexsym}

\theoremstyle{plain}

\def\R{{\mathbb R}}

\def\oH{\buildrel\circ\over H}
\def\oH1{\buildrel\circ\over H\kern-.02in{}^1}
\def\qed{{\hfill $\Box$}}

\begin{document}

\title{ Determination of the shape of the ear channel
   \thanks{key words:  acoustic waves, 
inverse problems, ear
    }
}

\author{
A.G. Ramm\\
Mathematics Department, 
Kansas State University, \\
 Manhattan, KS 66506-2602, USA\\
ramm@math.ksu.edu\\
}

\date{}

\maketitle\thispagestyle{empty}

\begin{abstract} 
It is proved that the measurement of the acoustic pressure 
on the ear membrane allows one to determine the shape of the ear channel 
uniquely.
\end{abstract}


\section{Introduction}
Consider a bounded domain $D \subset \R^n$, $n = 3,$ with a
Lipschitz  boundary $S$. Let $F$ be an open subset on $S$, a membrane, 
$G=S\setminus F$,  $\Gamma=\partial F$, and $N$ is the outer unit normal 
to $S$. 

Consider the problem:
$$\nabla^2 u+k^2 u=0 \hbox{\ in\ } D, \quad
  u = f \hbox{\ on\ } F,  \quad u = 0 \hbox{\ on\ } G. \eqno{(1.1)}$$
We assume that $k^2$ is not a Dirichlet eigenvalue of the
Laplacian in $D$. This assumption will be removed later.
If this assumption holds,  then the solution to problem
(1.1) is unique. Thus, its normal derivative, $h:=u_N$ on $F$, is uniquely 
determined. Suppose one can measure $h$ on $F$ for some $f\in C^1(F), 
\,\,f\not\equiv 0$.

The inverse problem (IP) we are interested in can now be formulated:
 
{\it Does this datum determine $G$ uniquely?}

Thus, we assume that $F$, $f$ and $h$ are known, that $k^2$ is not a 
Dirichlet eigenvalue of the Laplacian in $D$,  and want 
to determine the unknown part $G$ of the boundary $S$. 

Let $\Lambda$ be
the smallest eigenvalue of the Dirichlet Laplacian $L$ in $D$.
Let us assume that 
$$\Lambda>k^2.
\eqno{(1.2)}$$
Then, of course, problem (1.1) is uniquely solvable.
Assumption (1.2) in our problem is practically not a serious restriction, 
because the wavelength in our experiment can be chosen as we wish.
Since the upper bound on the width $d$ of the ear channel is known, 
and since 
$$\Lambda>\frac 1 { d^2}, \eqno{ (*)}$$ 
one can choose $k^2<\frac 
{1} {d^2}$ to satisfy assumption (1.2). A proof of the
estimate  (*) is given at the end of this note.

We discuss the Dirichlet condition but 
a similar argument is applicable to the Neumann and Robin boundary 
conditions. Boundary-value problems and scattering problems in rough 
domains were studied in [1].

Our basic result is the following theorem:
  
{\bf Theorem 1.} {\it If (1.2) holds then the above data determine $G$ 
uniquely}.

{\bf Remark 1.}  If $k^2$ is an eigenvalue of the Dirichlet
Laplacian $L$ in $D$, and $m(k)$ is the total multiplicity 
of the spectrum of $L$ on the semiaxis $\lambda\leq k^2$,
then $G$ is uniquely defined by the data $\{f_j, h_j\}_{1\leq j \leq 
m(k)+1}$, where $\{f_j\}_{1\leq j \leq m(k)+1}$ is an arbitrary fixed 
linearly independent system of functions in $C(F)$.

In Section 2 proofs are given.

\section{Proofs.}

{\bf Proof of Theorem 1.}   

Suppose that there are two surfaces $G_1$ and $G_2$, which generate the 
same data, that is, the same function $h$ on $F$.  Let $D_1, u_1$ and 
$D_2, u_2$ be the corresponding domains and 
solutions to (1.1). Denote $w:=u_1-u_2$, $D^{12}:=D_1\cap D_1$, 
$D_{12}:=D_1\cup D_2$, $D_3:=D_1\setminus D^{12}$, $D_4:=D_2\setminus 
D^{12}$. Note that $ w= w_N = 0$ on $F$, since the data 
$f$ and $h$ are the same by our assumption.

Threfore, one has:
$$\nabla^2 w+k^2w=0 \hbox{\ in\ } D^{12}, \quad
  w= w_N = 0 \hbox{\ on\ } F \eqno{(2.1)}$$
By the uniqueness of the solution to the Cauchy problem for elliptic 
equations, one concludes that $w=0$ in $ D^{12}$. Thus, $u_1=u_2=0$
on $\partial D^{12}$, and $u_1=0$ on $\partial D_3$. Thus
$$\nabla^2 u_1+k^2u_1=0 \hbox{\ in\ } D_3, \quad
  u_1 = 0 \hbox{\ on\ } \partial D_3. \eqno{(2.2)}$$
Since $D_3\subset D$, it follows that $\Lambda (D_3)>\Lambda (D)>k^2$.
Therefore $k^2$ is not a Dirichlet eigenvalue of the Laplacian in 
$D_3$, so $u_1=0$ in $D_3$, and, by the unique continuation property, 
$u_1=0$ in $D_1$. In particular, $u_1=0$ on $F$, which is a 
contradiction, since $u_1=f\neq 0$ on $F$ by the assumption. 
Theorem 1 is proved. \qed

{\bf Proof of Remark 1.} Suppose that $k^2>0$ is arbitrarily fixed, and 
the data are $\{f_j, h_j\}_{1\leq j\leq m(k)}$.
Using the same argument as in the proof of Theorem 1, one arrives at the 
conclusion (2.2) with $u_{j,1}$ in place of $u_1$, where $u_{j,1}$
solves (1.1) with $f=f_j$, $1\leq j \leq m(k)+1$.
Since the total multiplicity of the spectrum of the Dirichlet Laplacian 
in $D$ is not more that $m(k)$,  one can conclude that 
$D_1=D_2$. Remark 1 is proved. \qed

We do not discuss in this short note the possible methods for calculating 
$G$ from the data.  

{\it Proof of estimate (*)}. 

Let $\alpha$ be a unit vector, and
$d(\alpha)$ be the width of $D$ in the direction $\alpha$, that is, the 
distance between two planes, tangent to the boundary $S$ of $D$
and perpendicular to the vector $\alpha$, so that $D$ lies between these 
two planes. Let 
$$d:=\min_{\alpha}d(\alpha)>0.$$
By the variational definition of $\Lambda$ one has:
$\Lambda=\min \int_D|\nabla u|^2 dx,$ where the minimization is
taken over all $u\in H^1$, vanishing on $S$ and normalized, 
$||u||_{L^2(D)}=1$. Denote $s:=x_1, \, y:=(x_2,x_3),$
and choose the direction of $x_1-$axis along the direction $\alpha$,
which minimizes $d(\alpha)$, so that the width of $D$ in the direction 
of axis $x_1$ equals $d$. Then one has:
$$u(s,y)=\int_a^s u_t(t,y)dt,$$ so 
$$|u(s,y)|^2\leq \int_a^s |u_t(t,y)|^2dt 
(s-a)\leq d \int_a^b |u_t(t,y)|^2dt,$$ 
where $s=a$ and $s=b$ are the 
equations of the two tangent to $S$ planes, the distance between which is 
$d=b-a$, and $D$ is located between these planes.

Denote by $F_s$ the crossection of $D$ by the plane $x_1=s$, $a<s<b$.
Integrating the last inequality with respect to $y$ over $F_s$, 
and then with respect to $s$ between $a$ and $b$, one gets:
$$||u||^2_{L^2(D)}\leq d^2 ||\nabla u||^2,$$  
  which implies inequality (*). \qed

\end{document}